\documentclass[aps,prb,twocolumn,groupedaddress,showpacs]{revtex4}

\usepackage{amsmath}
\usepackage{bm}
\usepackage{graphicx}
\usepackage{hyperref}
\usepackage{color}
\usepackage{algorithm}
\usepackage{algorithmic}
\usepackage{amsfonts}
\usepackage{amssymb}
\usepackage{epsfig}

\usepackage[caption=false]{subfig}

\newcommand{\dcaplus}{DCA$^+\:$}
\newcommand{\phiplus}{$\phi^{(2)}\:$}
\newcommand{\bk}{{\bf k}}
\newcommand{\br}{{\bf r}}
\newcommand{\bX}{{\bf X}}
\newcommand{\bx}{{\bf x}}
\newcommand{\bK}{{\bf K}}

\newcommand{\wn}{{i\omega_n}}

\newcommand{\barG}{{\bar{G}}}

\emergencystretch=\maxdimen
\begin{document}

\title{Interlaced coarse-graining for the dynamic cluster approximation}
\author{P. Staar}
\affiliation{IBM Research -- Zurich, CH-8803 R\"uschlikon, Switzerland}
\author{M. Jiang}
\affiliation{Institute for Theoretical Physics, ETH Zurich, 8093 Zurich, Switzerland}
\author{U.R. H\"ahner}
\affiliation{Institute for Theoretical Physics, ETH Zurich, 8093 Zurich, Switzerland}
\author{T.A. Maier}
\affiliation{Computer Science and Mathematics Division and Center for Nanophase Materials Sciences, Oak Ridge National Laboratory, Oak Ridge, Tennessee 37831, USA}
\author{T.C.S. Schulthess}
\affiliation{Institute for Theoretical Physics, ETH Zurich, 8093 Zurich, Switzerland}
\date{\today }

\begin{abstract}
The dynamical cluster approximation (DCA) and its DCA$^+$ extension use
coarse-graining of the momentum space to reduce the complexity of quantum
many-body problems, thereby mapping the bulk lattice to a cluster embedded in
a dynamical mean-field host. Here, we introduce a new form of an interlaced
coarse-graining and compare it with the traditional coarse-graining. While it
gives a more localized self-energy for a given cluster size, we show that it
leads to more controlled results with weaker cluster shape and smoother
cluster size dependence, which converge to the results obtained from the
standard coarse-graining with increasing cluster size. Most importantly, the
new coarse-graining reduces the severity of the fermionic sign problem of the
underlying quantum Monte Carlo cluster solver and thus allows for calculations
on larger clusters. This enables the treatment of longer-ranged correlations
than those accessible with the standard coarse-graining and thus can allow for
the evaluation of the exact infinite cluster size result via finite size
scaling. As a demonstration, we study the hole-doped two-dimensional Hubbard
model and show that the interlaced coarse-graining in combination with the
extended DCA$^+$ algorithm permits the determination of the superconducting
$T_c$ on cluster sizes for which the results can be fit with a
Kosterlitz-Thouless scaling law.
\end{abstract}


\maketitle

\section{Introduction}

Much of the numerical work in the area of strongly correlated electron
materials is based on exact calculations that determine the state of a finite
size lattice and regard this state as an approximation of the thermodynamic
limit. The dynamic cluster approximation (DCA)
\cite{Hettler:1998wk,Maier:2005tj} uses a similar philosophy, in which the
bulk lattice problem is represented by a finite number of cluster degrees of
freedom. But in contrast to finite size calculations, the DCA uses
coarse-graining to retain information about the bulk degrees of freedom not
represented on the cluster. This leads to an approximation of the
thermodynamic limit, in which the bulk problem is replaced by a finite size
cluster embedded in a mean-field host that is designed to represent the rest
of the system. This approximation makes the problem tractable so it can be
solved with exact methods such as quantum Monte Carlo \cite{Jarrell:2001ty}.

To setup the cluster problem, one starts by dividing the first Brillouin zone
(BZ) into $N_c$ patches, each of which is represented by a cluster momentum $
{\bf K}$ (see Fig.~1, top left, for an example of a 16-site cluster)
\cite{Maier:2005tj}. One then assumes that the self-energy is well
approximated by a coarse-grained self-energy \cite{Maier:2005tj,Okamoto:2003vt}
\begin{equation} \label{eq:SigmaDCA}
	\Sigma^{\rm DCA}({\bf k},i\omega_n) = \sum_{\bf K}\phi_{\bf K}({\bf k})
	\Sigma_c ({\bf
	K},i\omega_n)\,.
\end{equation}
Here, $\Sigma_c(\bK,i\omega_n)$ is the self-energy of the $N_c$-site cluster
and the patch function $\phi_{\bf K}({\bf k}) = 1$ for ${\bf k}$ inside the
${\bf K}^{\rm th}$ patch, and 0 otherwise. One then coarse-graines the Green's
function
\begin{equation} \label{eq:cg}
	{\bar G}({\bf K},i\omega_n) = \frac{N_c}{N} \sum_{\bf k} \phi_{\bf K}({\bf k})
	\frac{1}{i\omega_n +\mu - \varepsilon_{\bf k}-\Sigma^{\rm DCA}({\bf k}
	,i\omega_n)}
\end{equation}
to set up an effective cluster problem, in which the cluster self-energy
${\Sigma}_c({\bf K},i\omega_n) = \Sigma_c[{{\cal G}_0({\bf K},i\omega_n)}]$ is
calculated as a functional of the corresponding bare cluster propagator $
{\cal G}_0({\bf K},i\omega_n) = [{\bar G}^{-1}({\bf K},i\omega_n)+\Sigma_c( {\bf
K},i\omega_n)]^{-1}$. The approximation in Eq.~(\ref{eq:SigmaDCA}) of the
lattice self-energy as a piecewise constant continuation of the cluster
self-energy leads to discontinuities between the patches and in some cases to
strong finite size effects, manifested as a strong dependence on the cluster
shape and size \cite{Staar:2013ec}.

In order to weaken these effects, the DCA method was recently extended through
the inclusion of a lattice self-energy with continuous momentum dependence
\cite{Staar:2013ec}.
This extended \dcaplus algorithm is obtained by reversing
Eq.~(\ref{eq:SigmaDCA}) to give a relation
\begin{equation}\label{eq:SigmaDCAP}
	\Sigma_c({\bf K},i\omega_n) = \frac{N_c}{N}\sum_{\bf k} \phi_{\bf K}({\bf k})
	\Sigma^{\rm DCA^+}({\bf k},i\omega_n)
\end{equation}
between the cluster self-energy $\Sigma_c({\bf K},i\omega_n)$ and the \dcaplus
lattice self-energy $\Sigma^{\rm DCA^+}({\bf k},i\omega_n)$. As discussed in
Ref.~\cite{Staar:2013ec}, $\Sigma^{\rm DCA^+}({\bf k},i\omega_n)$ with
continuous momentum dependence is then determined from a deconvolution of
Eq.~(\ref {eq:SigmaDCAP}) after the cluster self-energy $\Sigma_c({\bf
K},i\omega_n)$ is interpolated between the cluster ${\bf K}$ momenta. It was
shown that the \dcaplus algorithm reduces the cluster shape and size
dependence of the DCA self-energy, and, in addition, weakens the fermion sign
problem \cite{Troyer:2005ui} of the underlying QMC cluster solver.

One usually defines the patches as the Brillouin zones of the superlattice
(see Fig.~\ref{fig:phi_k} left) and sets $\phi_{\bf K}({\bf k})=1$ or 0 for
${\bf k}$ inside or outside the ${\bf K}^{th}$ patch, respectively
\cite{Maier:2005tj}. But the choice of coarse-graining patch functions
$\phi_{{\bf K}}({\bf k})$ in the DCA and \dcaplus is not unique. In
Ref.~\cite{Gull:2010by}, for example, Gull {\it et al.} used a star-like patch
geometry for a 4-site cluster to deform the central patch in order to capture
an important part of the Fermi surface. As we will discuss, there is a set of
constraints that must be satisfied by the patching. But these constraints
leave ample freedom in choosing different shapes of the coarse-graining
patches and different forms of the functions $\phi_{\bf K}({\bf k})$.

Here, we introduce a new interlaced coarse-graining, study its effects on the
self-energy of a single-band Hubbard model and compare the results with the
standard coarse-graining. For small cluster sizes, we find that the interlaced
coarse-graining leads to a more localized self-energy with less dependence on
the shape of the cluster. In the infinite cluster size limit, it gives results
that converge with those obtained from the standard coarse-graining. As an
important benefit, it significantly reduces the QMC fermion sign problem,
enabling calculations with larger cluster sizes. As an example, we show
results for the superconducting transition temperature, for which the
interlaced coarse-graining provides access to large enough clusters, so that
$T_c$ can be converged.

\section{Interlaced coarse-graining}

As noted, the patching must satisfy a number of constraints
\cite{Hettler:2000ud}. First, all patches must have equal size. This ensures
that the algebra of the operators of the effective cluster model obey the
usual fermionic algebra. Second, the patch functions should satisfy an
orthonormality condition, i.e. $\frac{N_c}{N}\sum_{\bf k}\phi_{\bf K}({\bf
k})\phi_{\bf K'}({\bf k}) =
\delta_{\bf KK'}$, so that different patches do not overlap or, in other
words, at any momentum ${\bf k}$, there is exactly one ${\bf K}$ for which
$\phi_{\bf K}({\bf k})$ is nonzero. Finally, the patches should have the same
symmetry as the cluster, so that the coarse-grained Green's function and
self-energy have the same symmetry as the cluster.

\begin{figure}[h!]
\begin{center}
\includegraphics[width=1.02\columnwidth]{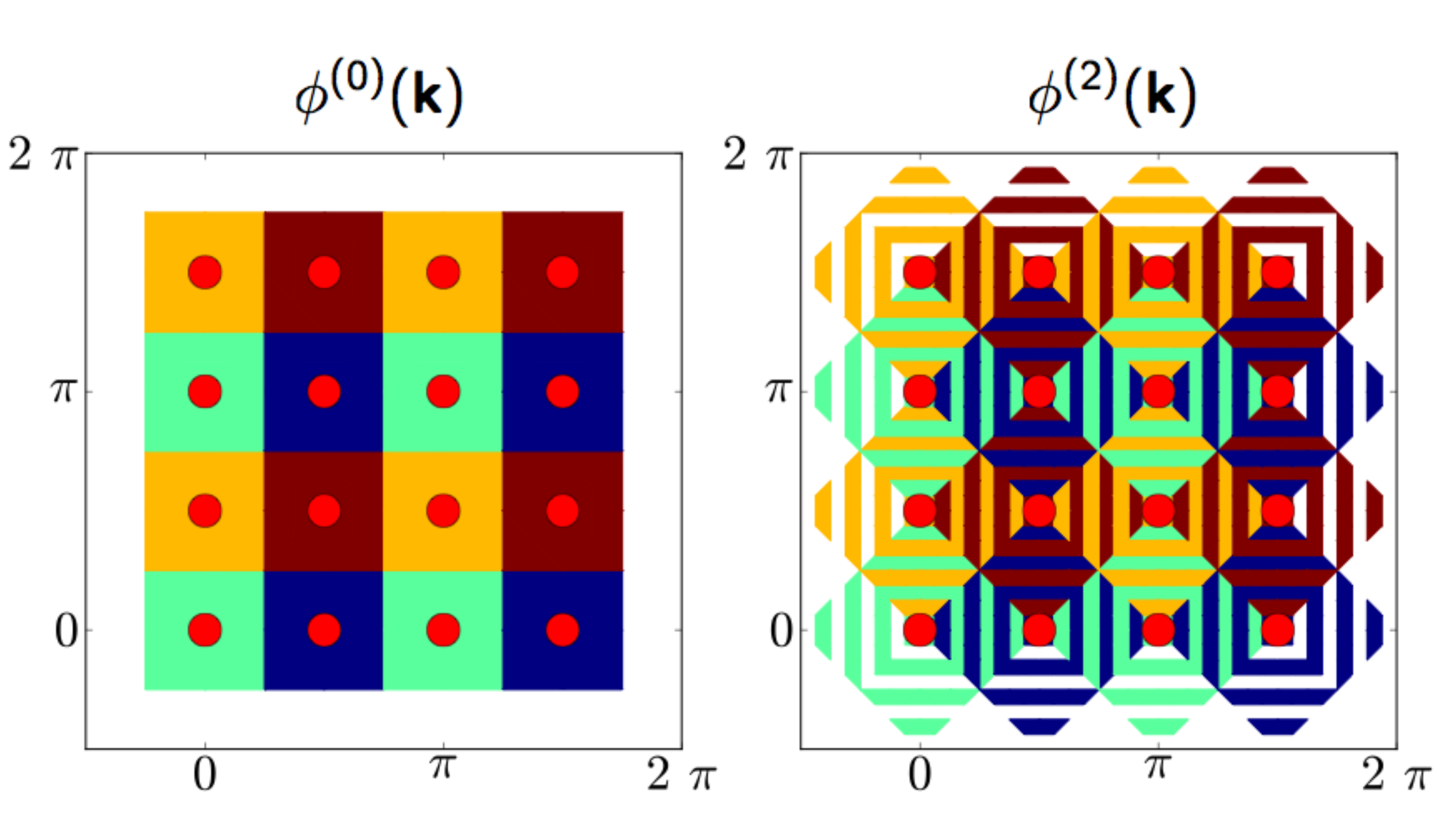}
\includegraphics[width=0.5\columnwidth]{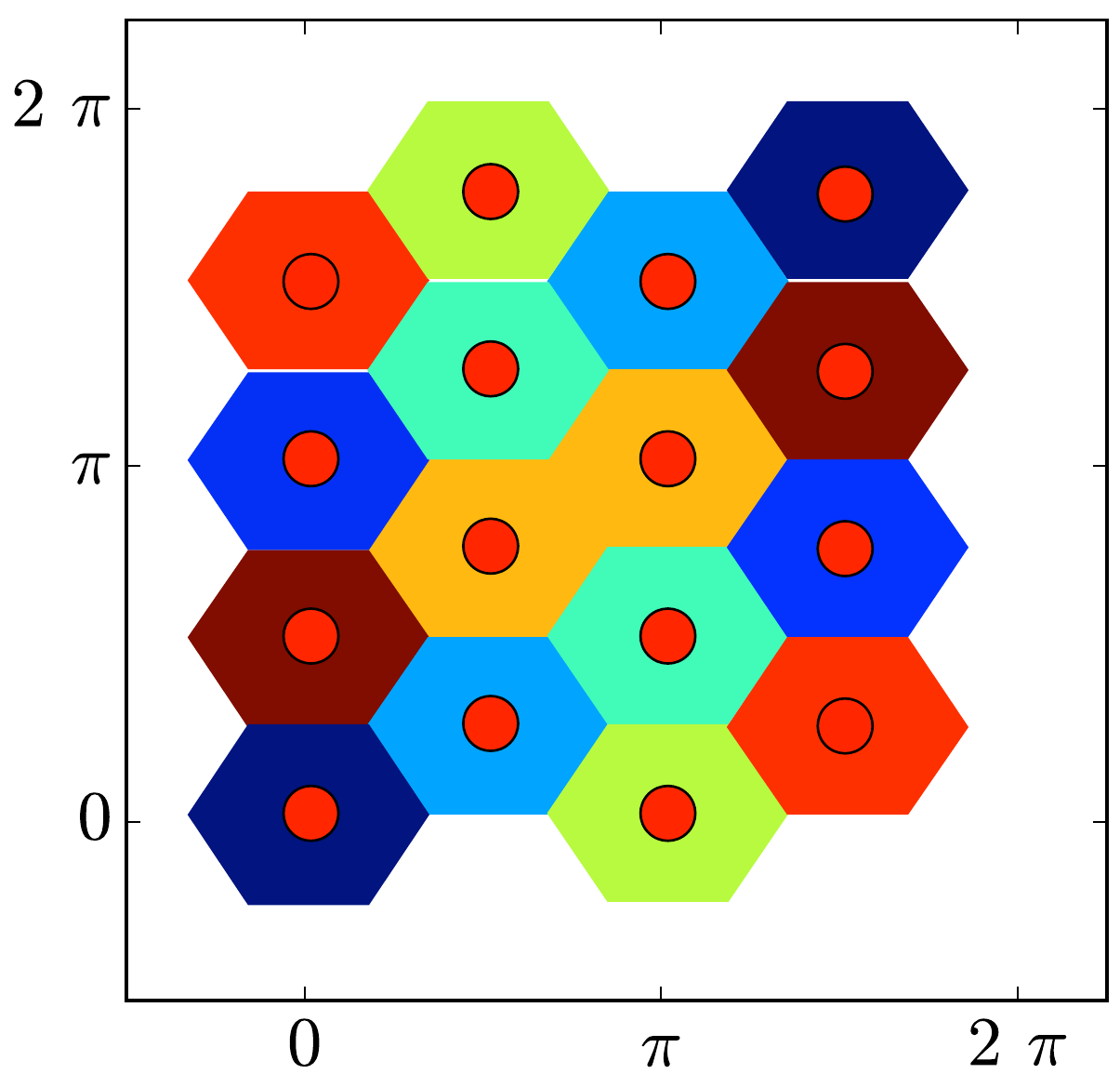}\includegraphics[width=0.5\columnwidth]{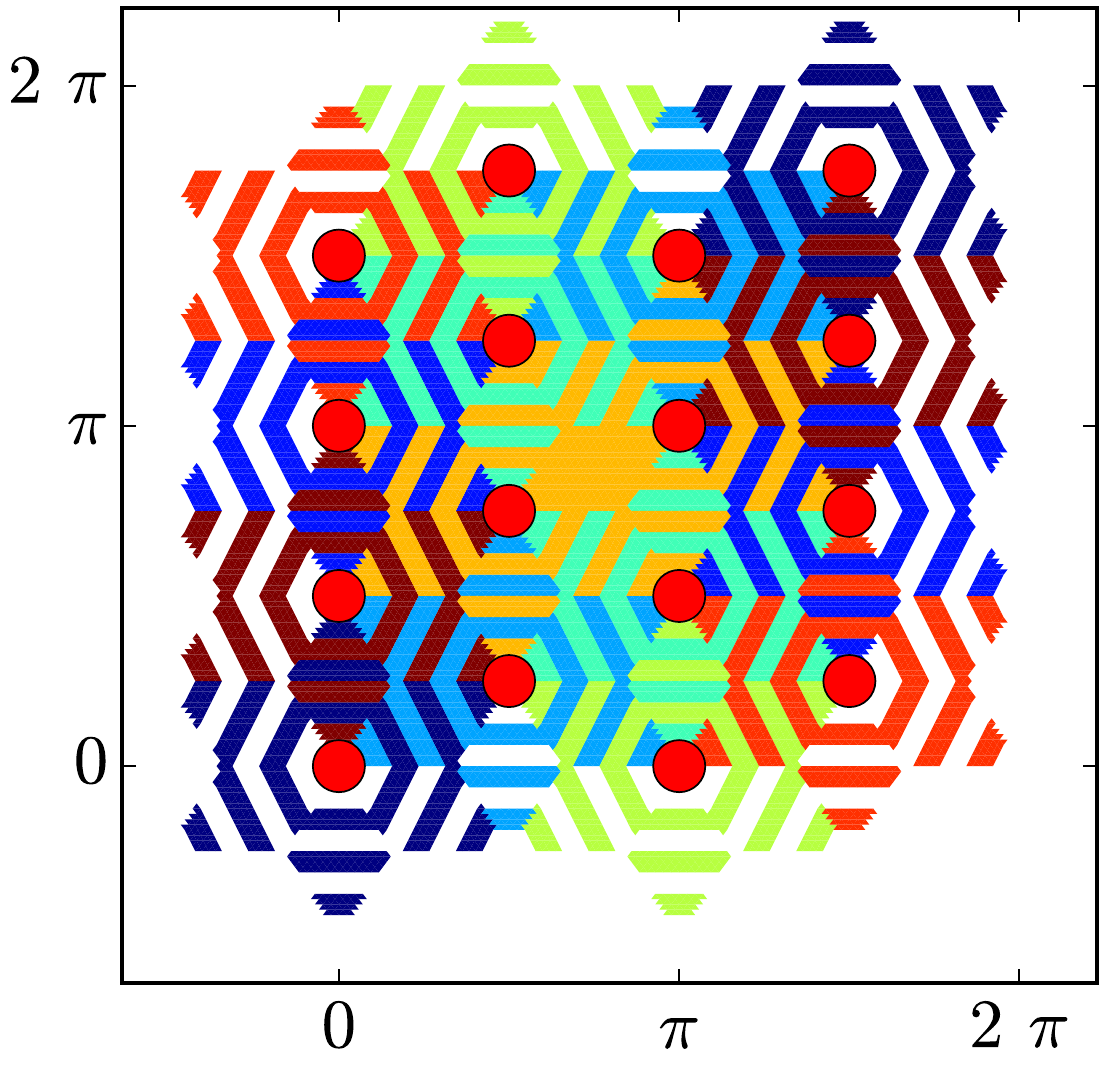}
\end{center}
\caption{\label{fig:phi_k} The location of the cluster momenta ${\bf K}$
and the shape of the patches for a 16A cluster (top) and 16B cluster (bottom).
The standard coarse-graining uses the Brillouin zone of the superlattice as the
patches (left), while the new coarse-graining uses patches, in which regions
assigned to neighboring ${\bf K}$ points are interleaved. }
\end{figure}

For the regular 4$\times$4 cluster labeled as 16A, the standard choice of the
coarse-graining patches, which we label $\phi^{(0)}({\bf k})$, is shown in
Fig.~\ref {fig:phi_k} in the top left panel. In the top right panel, we
introduce a new striped coarse-graining scheme defined by the patch functions
$\phi^{(2)}_{\bf K} ({\bf k})$, in which patches from neighboring $\bf
K$-points are interleaved. The generation of these patches is detailed in the
Appendix and the label $(2)$ indicates the number of stripes (see Appendix).
Obviously, these patches satisfy the constraints of equal volume,
orthonormality and symmetry. The bottom panels show the standard
coarse-graining $\phi^{(0)}$ and new patching $\phi^{(2)}$ for the case of
another 16-site cluster with different shape, the 16B cluster.

\begin{figure}[h!]
\begin{center}
\includegraphics{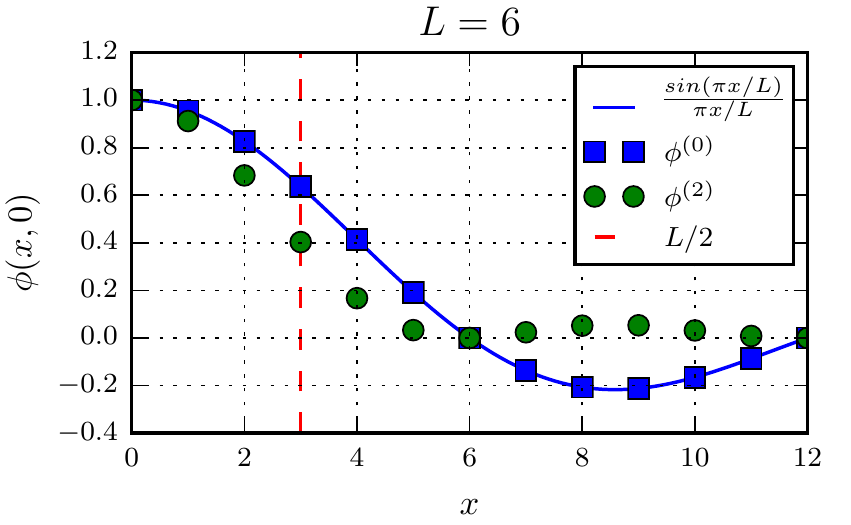}
\includegraphics{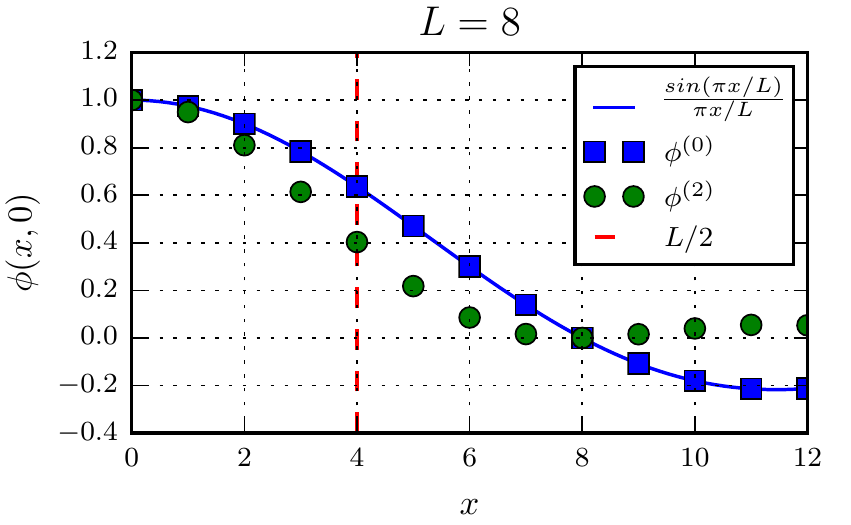}
\end{center}
\caption{\label{fig:phi_r} (Color online) Fourier-transform of the patch
functions $\phi^{(0)}({\bf k})$ and $\phi^{(2)}({\bf k})$ to real space for $L\times L$
clusters with $L=6$ (top) and $L=8$ (bottom) plotted versus $\br = (r,0)$.
$\phi^{(2)}(r)$ falls off more rapidly with $r$ than $\phi^{(0)}(r)$ and remains close
to 0 for $r\geq L$. }
\end{figure}

As discussed in Ref.~[\onlinecite{Staar:2013ec}], Eq.~(\ref{eq:cg}) may be
interpreted as a convolution of the lattice Green's function $G({\bf
k},i\omega_n)$ with the patch function $\phi_{\bf K}({\bf k})$, which may also
be written as $\phi({\bf k - K})$. Thus, the patch function essentially acts
as a filter. Since it is used to map the lattice problem onto the cluster
problem, it should pass the contribution to $G({\bf k},i\omega_n)$ that is
localized on the cluster, and cut off contributions outside the cluster.
Therefore, it is useful to investigate the coarse-graining in real space,
where, after Fourier-transforming Eq.~(\ref{eq:cg}), one has
\begin{equation}
	\label{eq:real_space_cg} {\bar G}({\bf X},i\omega_n) = \sum_{\bf
	x}\phi({\bf X+x}) G({\bf X+x},i\omega_n)\,.
\end{equation} Here, a vector ${\bf r=X+x}$ to a site in the real space bulk
lattice is broken up into a vector ${\bf X}$  within the real space cluster
and a vector ${\bf x}$ to the location of a cluster in the bulk lattice, and
$\phi ({\bf X+x})$ is the Fourier-transform to real space of the patch
function $\phi(\bf k)$. As noted by Hettler {\it et al.}~\cite
{Hettler:2000ud}, for a square cluster of size $L\times L$, one has for the
standard coarse-graining
\begin{equation}\label{eq:phir_sin}
	\phi^{(0)}({\bf r=X+x}) = \prod_{l=1}^2 \left[ \frac{\sin[\pi
	(x_l+X_l)/L]} {\pi(x_l+X_l)/L} \right]\,,
\end{equation}
where $x_l$ ($X_l$) is the $l^{\rm th}$ component of the vector ${\bf x}$ ($
{\bf X}$). Fig.~\ref{fig:phi_r} shows the ${\bf r}$ dependence of $\phi^{(0)}(
{\bf r})$ along the ${\bf x}$-direction ${\bf r} = (r,0)$ for an $L\times L$
cluster with $L=6$ (top) and $8$ (bottom). The sinusoidal dependence of
$\phi^{(0)}({\bf r})$ with the $1/r$ envelope from Eq.~(\ref{eq:phir_sin}) can
be seen. In the same figures, we also plot the $r$-dependence of the new
interlaced patch function $\phi^{(2)}(\br)$. One sees that $\phi^{(2)}(\br)$
falls off more rapidly with $r$ than the standard patch function
$\phi^{(0)}(\br)$. This may be understood from the fact that the $\phi^{(2)}$
coarse-graining averages over a more extended momentum region and thus leads
to a more local result. In addition, for distances $r > L$, $\phi^{(2)}(r)$
stays close to 0, while $\phi^{(0)}(\br)$ gives a significant negative
contribution to the coarse-grained average.  When the lattice Green's function
$G(\br)$ is short-ranged and drops to zero for $r\geq L/2$, only the $\bx
= 0$ term contributes to $\barG(\bX)$ in the coarse-graining sum in
Eq.~(\ref{eq:real_space_cg}), and hence $\barG (\bX) = \phi(\bX)G(\bX)$. In
this case, the standard $\phi^{(0)}$ coarse-graining gives a better
approximation, since $\phi^{(0)}(r)$ is closer to 1 for $r \leq L/2$ and thus
gives a $\barG(r)$ that is closer to the ``real'' $G(r)$. When $G(r)$ is
longer-ranged, however, the $\phi^{(0)}$ coarse-graining is less controlled,
since longer-ranged contributions from neighboring clusters can contribute
with either positive or negative weights, depending on the range $r$. This can
even lead to an overestimation of the short-range correlations within the
cluster. The $\phi^{(2)}$ coarse-graining, on the other hand, is always likely
to underestimate the non-local correlations and thus is more controlled. As
the cluster size increases, both approaches will give the same $\barG(r)$ once
$L/2$ is sufficiently large relative to the length-scale over which $G(r)$
vanishes.

\section{Application to the 2D Hubbard model}

Next we study the effects of these differences in the coarse-graining on the
momentum dependence of the self-energy. The Hubbard model that we study has a
nearest neighbor hopping $t$, a next-nearest neighbor hopping $t'$ and a
Coulomb repulsion $U$, and its Hamiltonian is
\begin{equation}\label{eq:Hubbard}
	H = \sum_{ij,\sigma} t_{ij} c^\dagger_{i\sigma}c^{\phantom\dagger}_
	{j\sigma} + U \sum_i n_{i\uparrow}n_{i\downarrow}\,.
\end{equation}
Here, $c^{ (\dagger)}_ {i\sigma}$ (creates) destroys an electron with spin
$\sigma$ on site $i$ and $n_{i\sigma} =
c^\dagger_{i\sigma}c^{\phantom\dagger}_{i\sigma}$ is the corresponding number
operator. To solve the effective cluster problem in the DCA and \dcaplus, we
use the continuous-time auxiliary-field QMC algorithm developed by Gull {\it
et al.} \cite{Gull:2008cm} with high-efficiency updates \cite{Gull:2011hh}.

\begin{figure}
\includegraphics{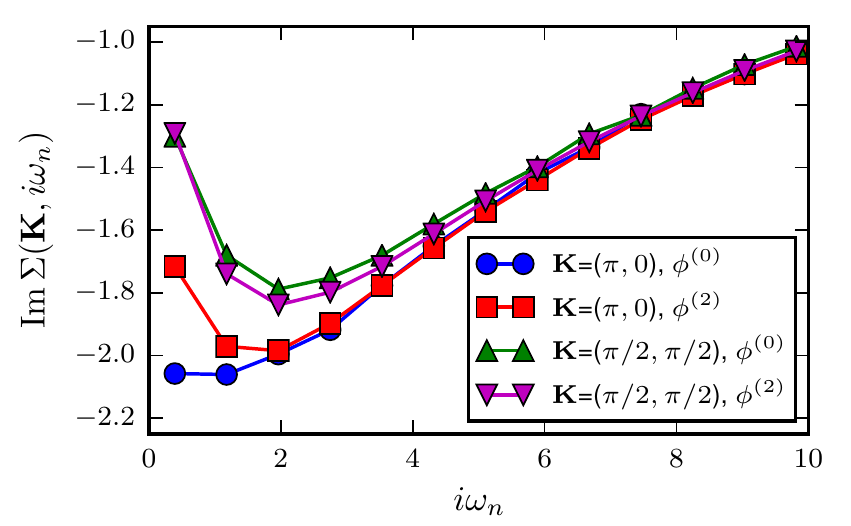}
\includegraphics{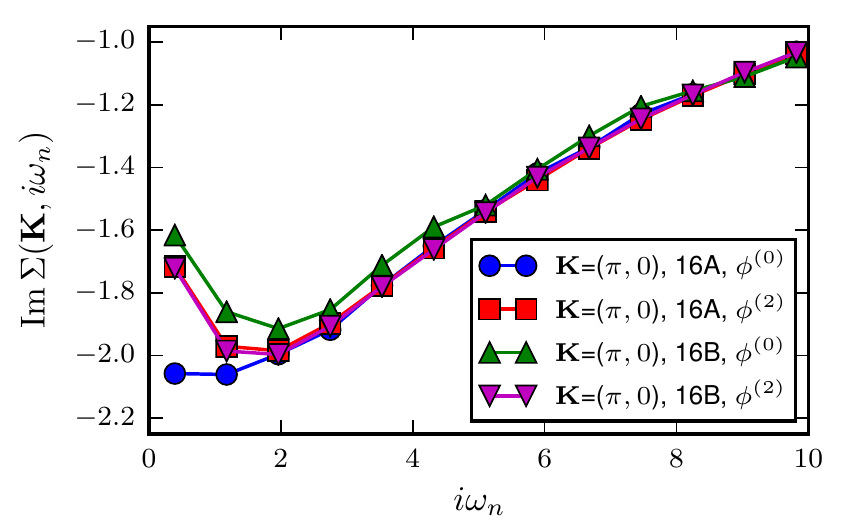}
\caption{\label{fig:Sigma} (Color online) Imaginary part of the self-energy for
different patching $\phi^{(0)}$ and $\phi^{(2)}$ for (a) the 16A cluster for ${\bf K}=
(\pi,0)$ and $(\pi/2,\pi/2)$ and (b) the 16A and 16B clusters for  ${\bf K}=
(\pi,0)$. The parameters are $t'=-0.15t$, $U=7t$, $\langle n \rangle = 0.942$
and $T=0.125$.}
\label{chis}
\end{figure}

The top panel of Fig.~\ref{fig:Sigma} shows results for the imaginary part of
the self-energy, ${\rm Im}\,\Sigma({\bf K},i\omega_n)$, for ${\bf K}= (\pi,0)$
and $(\pi/2,\pi/2)$ obtained for the 16A cluster. Here, we have set
$t'=-0.15t$, $U=7t$ and the site filling $\langle n \rangle=0.942$ and
temperature $T=0.125t$. For the standard coarse-graining $\phi^{(0)}$, one
observes a large difference in the low frequency behavior of ${\rm
Im}\,\Sigma({\bf K},i\omega_n)$ between ${\bf K}=(\pi,0)$ and $(\pi/2,\pi/2)$,
which has been observed in earlier DCA calculations (see e.g. the work in
Ref.~\cite{Gull:2010by}). The interlaced $\phi^{(2)}$ coarse-graining, in
contrast, gives a self-energy with much less momentum dependence. As expected
from the plots in Fig.~\ref{fig:phi_r} and their discussion, the $\phi^{(2)}$
patching gives a more local coarse-grained Green's function $\barG(r)$ and
thus a more local self-energy $\Sigma_c[\barG]$ with less momentum dependence.

The bottom panel of Fig.~\ref{fig:Sigma} shows results for ${\rm
Im}\,\Sigma({\bf K},i\omega_n)$ with $\bK=(\pi,0)$ for both the 16A and the 16B
clusters. Even though these clusters have the same size, the standard
$\phi^{(0)}$ coarse-graining gives results that vary significantly between the
two clusters, with qualitatively different behavior in the low frequency
region. In contrast, the \phiplus coarse-graining gives almost identical
results for these two cluster shapes. Again, this can be understood from the
fact that the interlaced \phiplus coarse-graining gives a more local
self-energy with weaker ${\bf k}$ dependence, which thus is less affected by
the location and shape of the coarse-graining patches.

\begin{figure}
\includegraphics{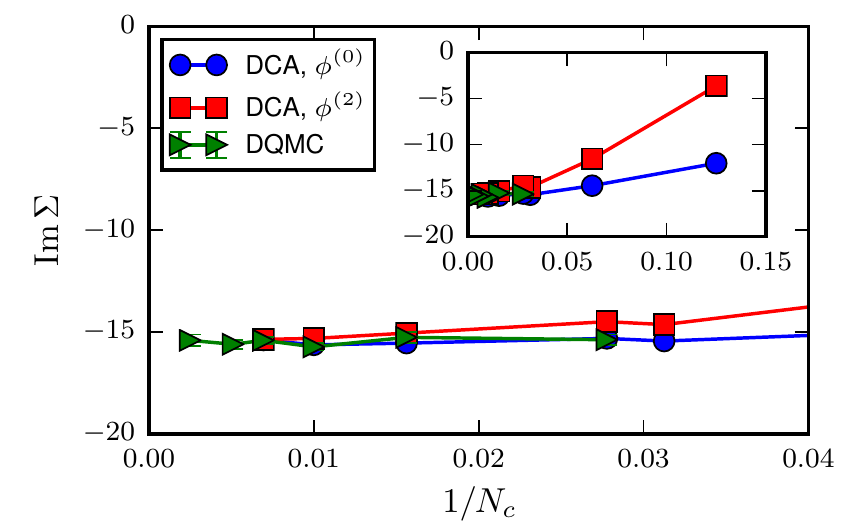}
\caption{\label{fig:Sigma2} (Color online) Imaginary part of the self-energy
for $\bK=(\pi,0)$ and $\omega_0=\pi T$ at half-filling $\langle n\rangle=1,
U=8t, T=0.15t$ versus the inverse cluster size $1/N_c$ obtained with DCA and
$\phi^{(0)}$ (blue) and \phiplus (red) coarse-graining and with determinantal
QMC calculations of a finite size lattice (black). With increasing cluster
size, the results obtained from the standard $\phi^{(0)}$ and the interlaced
$\phi^{(2)}$ coarse-graining converge to the same large cluster DQMC result.}
\end{figure}

As noted, one expects that this difference in the results from different forms
of the coarse-graining will decrease with increasing cluster size. To show
this, we plot in Fig.~\ref{fig:Sigma2} ${\rm Im}\, \Sigma(\bK,\pi T)$ with $\bK=
(\pi,0)$ for the half-filled $\langle n\rangle =1$ model with $t'=0$ and
$U=8t$ for the $\phi^{(0)}$ and $\phi^{(2)}$ patching. Also shown in this
figure are large cluster results obtained on a finite size lattice with the
determinantal QMC (DQMC) algorithm \cite{Blankenbecler:1981vf}. One sees that
for small cluster sizes, the standard $\phi^{(0)}$ coarse-graining gives much
better results that converge faster to the exact large $N_c$ limit, while the
$\phi^{(2)}$ coarse-graining underestimates the correlations. Again, this is
expected from the differences in real space $r$ behavior of the
$\phi^{(0)}(r)$ and $\phi^{ (2)}(r)$ shown in Fig.~\ref{fig:phi_r}. With
increasing cluster size, however, both curves converge to the same large
cluster DQMC result.

Next we turn to the effects of the coarse-graining on the fermion sign problem
of the underlying QMC solver. For the doped $\langle n\rangle \neq 1$ Hubbard
model in Eq.~(\ref {eq:Hubbard}), the sign problem is found to become
exponentially worse with increasing lattice size, decreasing temperature and
increasing $U$ \cite{Loh:1990hj}. QMC simulations of the doped model are
therefore limited to small lattices, high temperatures or weak coupling $U$.
QMC simulations within the framework of the DCA \cite
{Hettler:1998wk,Hettler:2000ud,Maier:2005tj} have been found to have a much
less severe sign problem than QMC simulations of finite size lattices
\cite{Jarrell:2001ty}. Lacking a rigorous mathematical argument, the DCA
improvement of the sign problem  was attributed to the action of the
mean-field host on the cluster \cite{Jarrell:2001ty}. This has allowed QMC
calculations at lower temperatures and larger $U$ than those
accessible by finite size QMC simulations \cite{Maier:2005tj}. Further
progress was made with the introduction of the \dcaplus method
\cite{Staar:2013ec}, which was found to exhibit an additional reduction of the
sign problem.  This was ascribed to the removal of artificial long-range
correlations, which arise in the DCA because of the discontinuities in the
self-energy, through a continuous lattice self-energy in the \dcaplus
\cite{Staar:2013ec}.

\begin{figure}[h!]
\begin{center}
\includegraphics{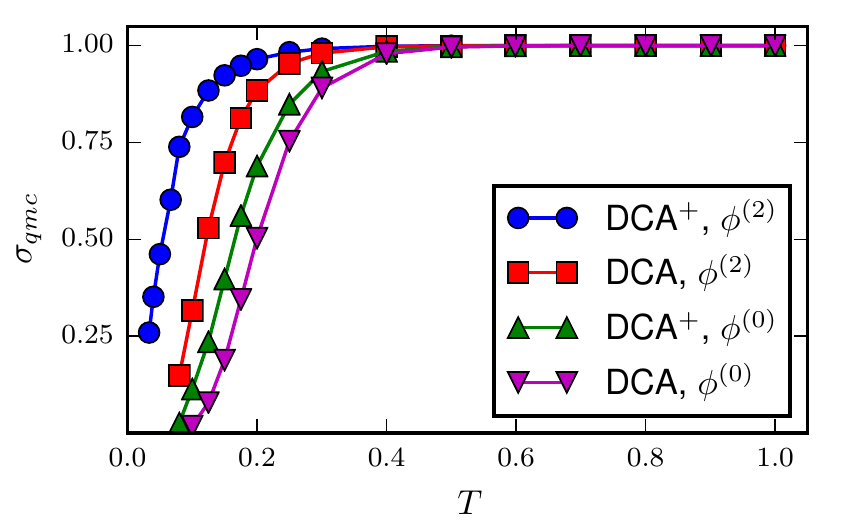}
\end{center}
\caption{\label{fig:fermionic_sign} The average QMC sign versus temperature for
$U/t=8$ and $\langle n\rangle = 0.9$ for the 16A  cluster. The use of the
\dcaplus and the $\phi^{(2)}$ patching lead to a significantly larger average
sign.}
\end{figure}

Here we study the effect of the coarse-graining on the sign problem.
Fig.~\ref{fig:fermionic_sign} shows the temperature dependence of the
average QMC sign for the 16A cluster with $t'=0$, $U=8t$ and $\langle n
\rangle = 0.9$ for both DCA and \dcaplus calculations with $\phi^{(0)}$ and
$\phi^{(2)}$ coarse-graining. At low temperatures, one sees that the QMC sign
falls rapidly to zero, and as noted, the \dcaplus algorithm gives an improved
sign relative to the DCA algorithm. As one sees, a significant further
improvement is achieved with the interlaced $\phi^{(2)}$ coarse-graining. When
combined with the \dcaplus algorithm, it has a significantly larger average
sign at low temperatures than the DCA algorithm with standard $\phi^{(0)}$
patching. For example, at $T=0.2t$, the sign in the DCA/$\phi^{(0)}$
calculation has fallen to 0.5, while the sign in the \dcaplus/$\phi^{(2)}$
calculation remains almost 1. As a consequence, one sees that the
\dcaplus/$\phi^{(2)}$ combination enables calculations with a sizeable sign at
much lower temperatures than those that are accessible with just the DCA or
the standard coarse-graining.

Finally, we illustrate the benefits of the improved sign problem by calculating
the superconducting transition temperature $T_c$ as a function of cluster
size $N_c$. To calculate $T_c$, we determine the eigenvalues $\lambda_\alpha$ and
eigenvectors $\phi_{\alpha}(k)$ of the Bethe-Salpeter equation \cite{Maier:2006ku}
\begin{equation}
	\label{eq:BSE}
	-\frac{T}{N}\sum_{k}\Gamma^{pp}(k,k')G(k')G(-k')\phi_\alpha
	(k')=\lambda_\alpha\phi_\alpha(k)\,,
\end{equation}
where $k=(\bk,\wn)$ and $\Gamma^{pp}(k,k')$ is the irreducible
particle-particle vertex on the bulk lattice. In the \dcaplus, just as the
self-energy in Eq.~(\ref{eq:SigmaDCAP}), the lattice vertex
$\Gamma^{pp}(k,k')$ is determined from inverting the equation
\begin{equation}
	\Gamma^{pp}_c(K,K') = \frac{N_c^2}{N^2}\sum_{\bK,\bK'} \phi_{\bK}(\bk)\Gamma^
	{pp}(k,k')\phi_{\bK'}(\bk')	
\end{equation}
as described in Ref.~\cite{Staar:2014jy}. At $T_c$, the leading eigenvalue
crosses 1 and one finds that the corresponding eigenvector has $d_ {x^2-y^2}$
symmetry \cite{Staar:2014jy}.

\begin{figure}
\begin{center}
\includegraphics{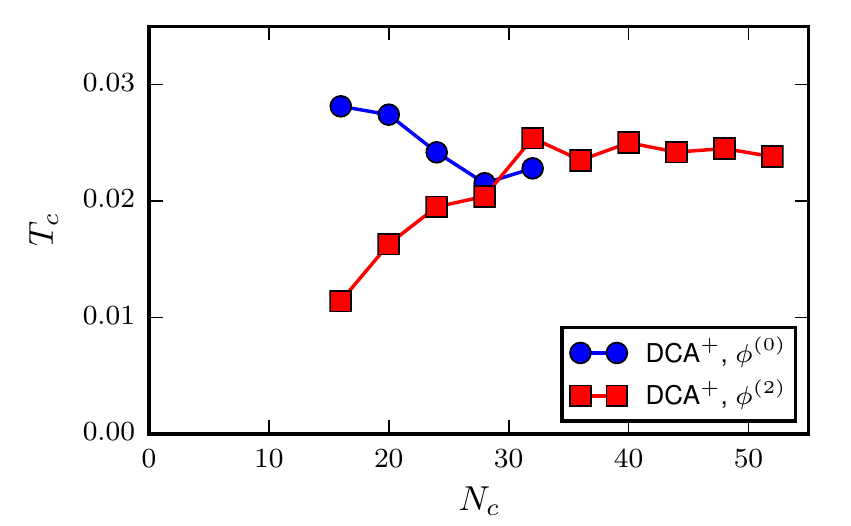}
\end{center}
\caption{\label{fig:Tc} (Color online) $d$-wave superconducting transition
temperature $T_c$ versus cluster size from \dcaplus calculations with the
interlaced $\phi^{(2)}$ and standard $\phi^{(0)}$ coarse-graining for $U=4t$, $t'=0$ and
$\langle n\rangle = 0.9$. The improved sign problem of the $\phi^{(2)}$
coarse-graining algorithm enables calculations on much larger clusters, for
which the results converge to the asymptotic large cluster limit.}
\end{figure}

In Fig.~\ref{fig:Tc}, we plot $T_c(N_c)$ for $U=4t$, $t'=0$ and $\langle
n\rangle=0.9$ from \dcaplus calculations with the standard $\phi^{(0)}$ and
the new interlaced $\phi^{(2)}$ coarse-graining. In the region with $N_c\leq
32$ the trends are clearly different: The $\phi^{(0)}$ coarse-graining gives a
$T_c$ that decreases with $N_c$ and thus apparently overestimates the pairing
correlations in small clusters, while the \phiplus coarse-graining gives an
increasing $T_c(N_c)$. This increase can again be traced to the stronger
locality of the \phiplus coarse-graining. The $d$-wave pairing strength arises
from a pairing interaction $\Gamma^{pp}(k,k')$ that increases
with momentum transfer $\bk-\bk'$ \cite{Maier:2006ku}. In small clusters, the
\phiplus coarse-graining underestimates this momentum dependence and thus
$T_c$. With increasing cluster size, this underestimation is reduced, and
$T_c$ increases with $N_c$. For $N_c=32$, both approaches give similar $T_c$.
While the sign problem of the standard coarse-graining prevents calculations
for $N_c>32$, the interlaced coarse-graining allows simulations of
significantly larger clusters. As one sees from Fig.~2, the $\phi^{(2)}$
coarse-graining takes into account correlations in these larger clusters that
have longer-range than those taken into account by the $\phi^{ (0)}$
coarse-graining in the smaller clusters. This is particularly useful for the
study of phase transitions where the critical behavior is determined by the
long-range correlations. One sees that the $\phi^{(2)}$ coarse-graining gives
results with smooth cluster size dependence in the $N_c \geq 32$ region, which
is not accessible by the standard $\phi^{(0)}$ coarse-graining. As previously
discussed in Ref.~\cite {Staar:2014jy}, it is possible to fit the results in
this region with a Kosterlitz-Thouless scaling curve and determine a
$T_c$ for the exact infinite cluster size limit.

\section{Summary and Conclusions}

To conclude, we have introduced and studied a new form of an interlaced
coarse-graining for the DCA and \dcaplus algorithms to map the bulk lattice to
an effective cluster problem and compared it with the standard
coarse-graining. This interlaced coarse-graining averages over a more extended
region in momentum space and thus gives a more localized self-energy with
weaker $k$-dependence. Non-local correlations are thus potentially
underestimated in small clusters and, in the absence of the QMC sign problem,
the standard coarse-graining converges faster to the exact infinite cluster
size result. However, the interlaced coarse-graining generally gives more
controlled results with weaker cluster shape and smoother cluster size
dependence that converge with the results from the standard coarse-graining
with increasing cluster size. Most importantly, we find that the interlaced
coarse-graining significantly reduces the sign problem of the underlying QMC
solver, thereby enabling calculations with larger cluster sizes, for which
longer-ranged correlations are taken into account and the underestimation of
shorter-ranged correlations is not an issue. The new coarse-graining is thus
particularly well suited for large cluster studies of phase transitions where
the critical behavior is determined by the long-range correlations. As an
example, we have shown that the interlaced coarse-graining in combination with
the \dcaplus algorithm enables calculations of the superconducting $T_c$ on
cluster sizes, for which the results converge to the asymptotic
Kosterlitz-Thouless scaling curve. Thus, while care should be taken in
interpreting results on small clusters, the new coarse-graining introduced in
this paper gives access to much larger cluster sizes and thus can enable a
finite size scaling analysis to recover the exact infinite cluster size
result.

\begin{acknowledgments}
We would like to thank Richard Scalettar for useful comments. Part of this
research was conducted at the Center for Nanophase Materials Sciences, which
is a DOE Office of Science User Facility. An award of computer time was
provided by the Innovative and Novel Computational Impact on Theory and
Experiment (INCITE) program. This research used resources of the Oak Ridge
Leadership Computing Facility, which is a DOE Office of Science User Facility
supported under Contract DE-AC05-00OR22725.
\end{acknowledgments}

\appendix*
\section{Setup of interlaced coarse-graining patches}

\begin{figure}[h!]
	\begin{center}
		\includegraphics[width=14cm, trim = 85mm 35mm 0mm 20mm, clip]{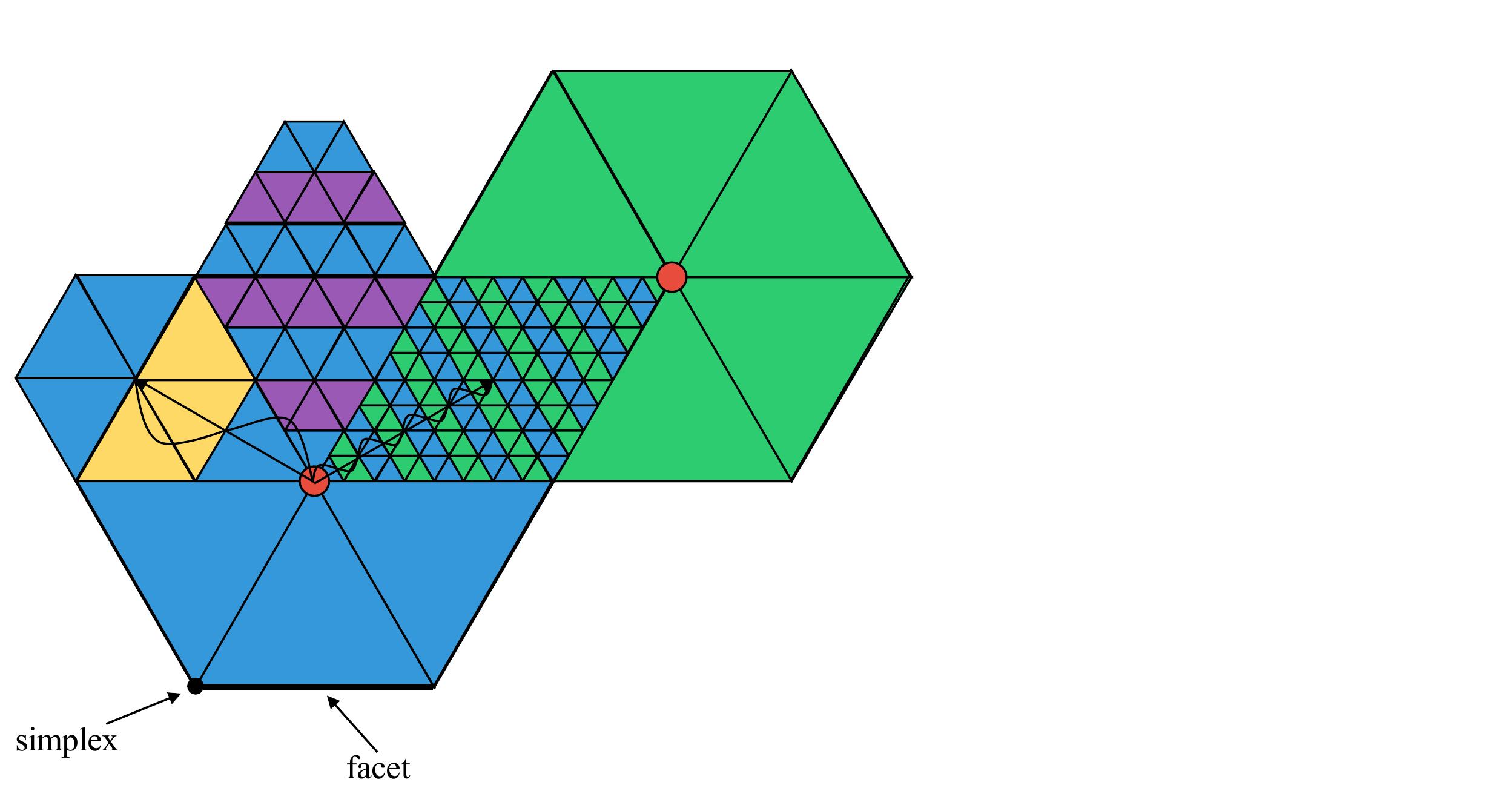}
	\end{center}
	\caption{\label{fig:patch_construction} (Color online) Construction of the
	patches for the 16B cluster. Successive recursion steps of the k-mesh
	refinement are shown: 0 (plain blue), 1 (blue-yellow), 2 (blue-purple) and
	3 (blue-green). Stripes are reflected with the maximum number of periods
	possible: 0, 1, 2 and 4. The sine, that determines whether a stripe is
	reflected, is sketched for $p=1$ and $p=4$ in recursion step 1 and 3,
	respectively.}
\end{figure}

In this section we describe the algorithm that generates the interlaced
coarse-graining patches for the two dimensional case. We start from the
traditional coarse-graining patches defined as the Brillouin zones of the
superlattice. We call the edges and corners of these patches \emph{facets} and
\emph{simplexes}, respectively. First, each initial patch is divided into
triangles by connecting the corners of the patch with its center, i.e. the
cluster momentum $\bK$. For example, the square shaped initial patches of the
16A cluster are split into four triangles while the hexagonal shaped initial
patches of the 16B cluster are divided into six triangles. Each of these
triangles has a unique facet as one edge. The vector perpendicular to the
facet that connects the center to the facet is defined as the normal vector
$\bf{n}$ of the triangle. We then recursively subdivide each triangle into
four similar triangles by connecting the midpoints of each side. By
construction, this \emph{k-mesh refinement} at the same time divides the
initial triangle into stripes of equal width parallel to its facet (see
Fig.~\ref{fig:patch_construction}). The details of the first three recursion
steps are listed in Tab.~\ref{tab:k-mesh-refinement_16B} for the 16B cluster.

\begin{table}[h]
	\centering
	\begin{tabular}{ccccccc}
		\hline \hline
		recursion & 0 & 1 & 2 & 3 && k\\ \hline
		no. triangles & 6 & 24 & 96 & 384 && $6\times4^k$ \\
		no. stripes & 1 & 2 & 4 & 8 &&  $2^k$ \\
		max. no. periods & 0 & 1 & 2 & 4 && $2^{k-1}$ \\
		\hline \hline
	\end{tabular}
	\caption{Details of the k-mesh refinement for the 16B cluster up to three
	recursion steps. The initial number of triangles for the hexagonal shaped
	patch is six. Each recursion step divides the triangles into four similar,
	smaller triangles. The number of stripes is doubled each time. The maximum
	number of periods is the number of stripes divided by two.}
	\label{tab:k-mesh-refinement_16B}
\end{table}

For each of the smallest triangles we compute the center of mass $\bf{k}_\mathrm{cm}$ and project it onto the normal vector $\bf{n}$ of its initial triangle. The interlaced patches of \emph{period $p$} are then obtained by reflecting those triangles across this facet for which
\begin{equation}
\sin\left( \frac{2 \pi \, p}{{\bf n} \cdot {\bf n}} \, \bf{k}_\mathrm{cm} \cdot \bf{n} \right) < 0 \,.
\end{equation} For the 16B cluster this is illustrated in
Fig.~\ref{fig:patch_construction} for one period (blue-yellow) and three
periods (blue-green).\\ First note that by construction stripes always get
reflected as a whole. As its name indicates, $p$ is just the number of periods
of the sine along the normal vector $\bf{n}$ and as such twice the numbers of
reflected stripes. Therefore, the total number of stripes in the initial
triangle must be a multiple of the number of periods, $p$, and at least $2p$.
Consequently, the number of periods determines the minimum number of recursion
steps required.

One advantage of this new approach of generating the coarse-graining patches
is its recursive nature. The larger the recursion depth, the more the patches
become interlaced. But at the same time the only geometric structures
occurring are triangles, which are easy to integrate over. The traditional
coarse-graining is a special case and corresponds to zero recursion steps.
Last but not least, this new coarse-graining approach can easily be
generalized to three dimensions, in which triangles become tetrahedrons.

\bibliography{refs.bib}

\begin{thebibliography}{14}%
\makeatletter
\providecommand \@ifxundefined [1]{%
 \@ifx{#1\undefined}
}%
\providecommand \@ifnum [1]{%
 \ifnum #1\expandafter \@firstoftwo
 \else \expandafter \@secondoftwo
 \fi
}%
\providecommand \@ifx [1]{%
 \ifx #1\expandafter \@firstoftwo
 \else \expandafter \@secondoftwo
 \fi
}%
\providecommand \natexlab [1]{#1}%
\providecommand \enquote  [1]{``#1''}%
\providecommand \bibnamefont  [1]{#1}%
\providecommand \bibfnamefont [1]{#1}%
\providecommand \citenamefont [1]{#1}%
\providecommand \href@noop [0]{\@secondoftwo}%
\providecommand \href [0]{\begingroup \@sanitize@url \@href}%
\providecommand \@href[1]{\@@startlink{#1}\@@href}%
\providecommand \@@href[1]{\endgroup#1\@@endlink}%
\providecommand \@sanitize@url [0]{\catcode `\\12\catcode `\$12\catcode
  `\&12\catcode `\#12\catcode `\^12\catcode `\_12\catcode `\%12\relax}%
\providecommand \@@startlink[1]{}%
\providecommand \@@endlink[0]{}%
\providecommand \url  [0]{\begingroup\@sanitize@url \@url }%
\providecommand \@url [1]{\endgroup\@href {#1}{\urlprefix }}%
\providecommand \urlprefix  [0]{URL }%
\providecommand \Eprint [0]{\href }%
\providecommand \doibase [0]{http://dx.doi.org/}%
\providecommand \selectlanguage [0]{\@gobble}%
\providecommand \bibinfo  [0]{\@secondoftwo}%
\providecommand \bibfield  [0]{\@secondoftwo}%
\providecommand \translation [1]{[#1]}%
\providecommand \BibitemOpen [0]{}%
\providecommand \bibitemStop [0]{}%
\providecommand \bibitemNoStop [0]{.\EOS\space}%
\providecommand \EOS [0]{\spacefactor3000\relax}%
\providecommand \BibitemShut  [1]{\csname bibitem#1\endcsname}%
\let\auto@bib@innerbib\@empty
\bibitem [{\citenamefont {Hettler}\ \emph {et~al.}(1998)\citenamefont
  {Hettler}, \citenamefont {Tahvildar-Zadeh}, \citenamefont {Jarrell},
  \citenamefont {Pruschke},\ and\ \citenamefont
  {Krishnamurthy}}]{Hettler:1998wk}%
  \BibitemOpen
  \bibfield  {author} {\bibinfo {author} {\bibfnamefont {M.~H.}\ \bibnamefont
  {Hettler}}, \bibinfo {author} {\bibfnamefont {A.~N.}\ \bibnamefont
  {Tahvildar-Zadeh}}, \bibinfo {author} {\bibfnamefont {M.}~\bibnamefont
  {Jarrell}}, \bibinfo {author} {\bibfnamefont {T.}~\bibnamefont {Pruschke}}, \
  and\ \bibinfo {author} {\bibfnamefont {H.~R.}\ \bibnamefont
  {Krishnamurthy}},\ }\href@noop {} {\bibfield  {journal} {\bibinfo  {journal}
  {Physical Review B}\ }\textbf {\bibinfo {volume} {58}},\ \bibinfo {pages}
  {R7475} (\bibinfo {year} {1998})}\BibitemShut {NoStop}%
\bibitem [{\citenamefont {Maier}\ \emph {et~al.}(2005)\citenamefont {Maier},
  \citenamefont {Jarrell}, \citenamefont {Pruschke},\ and\ \citenamefont
  {Hettler}}]{Maier:2005tj}%
  \BibitemOpen
  \bibfield  {author} {\bibinfo {author} {\bibfnamefont {T.}~\bibnamefont
  {Maier}}, \bibinfo {author} {\bibfnamefont {M.}~\bibnamefont {Jarrell}},
  \bibinfo {author} {\bibfnamefont {T.}~\bibnamefont {Pruschke}}, \ and\
  \bibinfo {author} {\bibfnamefont {M.}~\bibnamefont {Hettler}},\ }\href@noop
  {} {\bibfield  {journal} {\bibinfo  {journal} {Reviews of Modern Physics}\
  }\textbf {\bibinfo {volume} {77}},\ \bibinfo {pages} {1027} (\bibinfo {year}
  {2005})}\BibitemShut {NoStop}%
\bibitem [{\citenamefont {Jarrell}\ \emph {et~al.}(2001)\citenamefont
  {Jarrell}, \citenamefont {Maier}, \citenamefont {Huscroft},\ and\
  \citenamefont {Moukouri}}]{Jarrell:2001ty}%
  \BibitemOpen
  \bibfield  {author} {\bibinfo {author} {\bibfnamefont {M.}~\bibnamefont
  {Jarrell}}, \bibinfo {author} {\bibfnamefont {T.}~\bibnamefont {Maier}},
  \bibinfo {author} {\bibfnamefont {C.}~\bibnamefont {Huscroft}}, \ and\
  \bibinfo {author} {\bibfnamefont {S.}~\bibnamefont {Moukouri}},\ }\href@noop
  {} {\bibfield  {journal} {\bibinfo  {journal} {Physical Review B}\ }\textbf
  {\bibinfo {volume} {64}},\ \bibinfo {pages} {195130} (\bibinfo {year}
  {2001})}\BibitemShut {NoStop}%
\bibitem [{\citenamefont {Okamoto}\ \emph {et~al.}(2003)\citenamefont
  {Okamoto}, \citenamefont {Millis}, \citenamefont {Monien},\ and\
  \citenamefont {Fuhrmann}}]{Okamoto:2003vt}%
  \BibitemOpen
  \bibfield  {author} {\bibinfo {author} {\bibfnamefont {S.}~\bibnamefont
  {Okamoto}}, \bibinfo {author} {\bibfnamefont {A.~J.}\ \bibnamefont {Millis}},
  \bibinfo {author} {\bibfnamefont {H.}~\bibnamefont {Monien}}, \ and\ \bibinfo
  {author} {\bibfnamefont {A.}~\bibnamefont {Fuhrmann}},\ }\href@noop {}
  {\bibfield  {journal} {\bibinfo  {journal} {Physical Review B}\ }\textbf
  {\bibinfo {volume} {68}},\ \bibinfo {pages} {195121} (\bibinfo {year}
  {2003})}\BibitemShut {NoStop}%
\bibitem [{\citenamefont {Staar}\ \emph {et~al.}(2013)\citenamefont {Staar},
  \citenamefont {Maier},\ and\ \citenamefont {Schulthess}}]{Staar:2013ec}%
  \BibitemOpen
  \bibfield  {author} {\bibinfo {author} {\bibfnamefont {P.}~\bibnamefont
  {Staar}}, \bibinfo {author} {\bibfnamefont {T.}~\bibnamefont {Maier}}, \ and\
  \bibinfo {author} {\bibfnamefont {T.~C.}\ \bibnamefont {Schulthess}},\
  }\href@noop {} {\bibfield  {journal} {\bibinfo  {journal} {Physical Review
  B}\ }\textbf {\bibinfo {volume} {88}},\ \bibinfo {pages} {115101} (\bibinfo
  {year} {2013})}\BibitemShut {NoStop}%
\bibitem [{\citenamefont {Troyer}\ and\ \citenamefont
  {Wiese}(2005)}]{Troyer:2005ui}%
  \BibitemOpen
  \bibfield  {author} {\bibinfo {author} {\bibfnamefont {M.}~\bibnamefont
  {Troyer}}\ and\ \bibinfo {author} {\bibfnamefont {U.}~\bibnamefont {Wiese}},\
  }\href@noop {} {\bibfield  {journal} {\bibinfo  {journal} {Physical Review
  Letters}\ }\textbf {\bibinfo {volume} {94}},\ \bibinfo {pages} {170201}
  (\bibinfo {year} {2005})}\BibitemShut {NoStop}%
\bibitem [{\citenamefont {Gull}\ \emph {et~al.}(2010)\citenamefont {Gull},
  \citenamefont {Ferrero}, \citenamefont {Parcollet}, \citenamefont {Georges},\
  and\ \citenamefont {Millis}}]{Gull:2010by}%
  \BibitemOpen
  \bibfield  {author} {\bibinfo {author} {\bibfnamefont {E.}~\bibnamefont
  {Gull}}, \bibinfo {author} {\bibfnamefont {M.}~\bibnamefont {Ferrero}},
  \bibinfo {author} {\bibfnamefont {O.}~\bibnamefont {Parcollet}}, \bibinfo
  {author} {\bibfnamefont {A.}~\bibnamefont {Georges}}, \ and\ \bibinfo
  {author} {\bibfnamefont {A.}~\bibnamefont {Millis}},\ }\href@noop {}
  {\bibfield  {journal} {\bibinfo  {journal} {Physical Review B}\ }\textbf
  {\bibinfo {volume} {82}},\ \bibinfo {pages} {155101} (\bibinfo {year}
  {2010})}\BibitemShut {NoStop}%
\bibitem [{\citenamefont {Hettler}\ \emph {et~al.}(2000)\citenamefont
  {Hettler}, \citenamefont {Mukherjee}, \citenamefont {Jarrell},\ and\
  \citenamefont {Krishnamurthy}}]{Hettler:2000ud}%
  \BibitemOpen
  \bibfield  {author} {\bibinfo {author} {\bibfnamefont {M.}~\bibnamefont
  {Hettler}}, \bibinfo {author} {\bibfnamefont {M.}~\bibnamefont {Mukherjee}},
  \bibinfo {author} {\bibfnamefont {M.}~\bibnamefont {Jarrell}}, \ and\
  \bibinfo {author} {\bibfnamefont {H.}~\bibnamefont {Krishnamurthy}},\
  }\href@noop {} {\bibfield  {journal} {\bibinfo  {journal} {Physical Review
  B}\ }\textbf {\bibinfo {volume} {61}},\ \bibinfo {pages} {12739} (\bibinfo
  {year} {2000})}\BibitemShut {NoStop}%
\bibitem [{\citenamefont {Gull}\ \emph {et~al.}(2008)\citenamefont {Gull},
  \citenamefont {Werner}, \citenamefont {Parcollet},\ and\ \citenamefont
  {Troyer}}]{Gull:2008cm}%
  \BibitemOpen
  \bibfield  {author} {\bibinfo {author} {\bibfnamefont {E.}~\bibnamefont
  {Gull}}, \bibinfo {author} {\bibfnamefont {P.}~\bibnamefont {Werner}},
  \bibinfo {author} {\bibfnamefont {O.}~\bibnamefont {Parcollet}}, \ and\
  \bibinfo {author} {\bibfnamefont {M.}~\bibnamefont {Troyer}},\ }\href@noop {}
  {\bibfield  {journal} {\bibinfo  {journal} {Europhysics Letters}\ }\textbf
  {\bibinfo {volume} {82}},\ \bibinfo {pages} {57003} (\bibinfo {year}
  {2008})}\BibitemShut {NoStop}%
\bibitem [{\citenamefont {Gull}\ \emph {et~al.}(2011)\citenamefont {Gull},
  \citenamefont {Staar}, \citenamefont {Fuchs}, \citenamefont {Nukala},
  \citenamefont {Summers}, \citenamefont {Pruschke}, \citenamefont
  {Schulthess},\ and\ \citenamefont {Maier}}]{Gull:2011hh}%
  \BibitemOpen
  \bibfield  {author} {\bibinfo {author} {\bibfnamefont {E.}~\bibnamefont
  {Gull}}, \bibinfo {author} {\bibfnamefont {P.}~\bibnamefont {Staar}},
  \bibinfo {author} {\bibfnamefont {S.}~\bibnamefont {Fuchs}}, \bibinfo
  {author} {\bibfnamefont {P.}~\bibnamefont {Nukala}}, \bibinfo {author}
  {\bibfnamefont {M.}~\bibnamefont {Summers}}, \bibinfo {author} {\bibfnamefont
  {T.}~\bibnamefont {Pruschke}}, \bibinfo {author} {\bibfnamefont
  {T.}~\bibnamefont {Schulthess}}, \ and\ \bibinfo {author} {\bibfnamefont
  {T.}~\bibnamefont {Maier}},\ }\href@noop {} {\bibfield  {journal} {\bibinfo
  {journal} {Physical Review B}\ }\textbf {\bibinfo {volume} {83}},\ \bibinfo
  {pages} {075122} (\bibinfo {year} {2011})}\BibitemShut {NoStop}%
\bibitem [{\citenamefont {Blankenbecler}\ \emph {et~al.}(1981)\citenamefont
  {Blankenbecler}, \citenamefont {Scalapino},\ and\ \citenamefont
  {Sugar}}]{Blankenbecler:1981vf}%
  \BibitemOpen
  \bibfield  {author} {\bibinfo {author} {\bibfnamefont {R.}~\bibnamefont
  {Blankenbecler}}, \bibinfo {author} {\bibfnamefont {D.}~\bibnamefont
  {Scalapino}}, \ and\ \bibinfo {author} {\bibfnamefont {R.}~\bibnamefont
  {Sugar}},\ }\href@noop {} {\bibfield  {journal} {\bibinfo  {journal}
  {Physical Review D}\ }\textbf {\bibinfo {volume} {24}},\ \bibinfo {pages}
  {2278} (\bibinfo {year} {1981})}\BibitemShut {NoStop}%
\bibitem [{\citenamefont {Loh}\ \emph {et~al.}(1990)\citenamefont {Loh},
  \citenamefont {Gubernatis}, \citenamefont {Scalettar}, \citenamefont {White},
  \citenamefont {Scalapino},\ and\ \citenamefont {Sugar}}]{Loh:1990hj}%
  \BibitemOpen
  \bibfield  {author} {\bibinfo {author} {\bibfnamefont {E.~Y.}\ \bibnamefont
  {Loh}}, \bibinfo {author} {\bibfnamefont {J.~E.}\ \bibnamefont {Gubernatis}},
  \bibinfo {author} {\bibfnamefont {R.~T.}\ \bibnamefont {Scalettar}}, \bibinfo
  {author} {\bibfnamefont {S.~R.}\ \bibnamefont {White}}, \bibinfo {author}
  {\bibfnamefont {D.~J.}\ \bibnamefont {Scalapino}}, \ and\ \bibinfo {author}
  {\bibfnamefont {R.~L.}\ \bibnamefont {Sugar}},\ }\href@noop {} {\bibfield
  {journal} {\bibinfo  {journal} {Physical Review B}\ }\textbf {\bibinfo
  {volume} {41}},\ \bibinfo {pages} {9301} (\bibinfo {year}
  {1990})}\BibitemShut {NoStop}%
\bibitem [{\citenamefont {Maier}\ \emph {et~al.}(2006)\citenamefont {Maier},
  \citenamefont {Jarrell},\ and\ \citenamefont {Scalapino}}]{Maier:2006ku}%
  \BibitemOpen
  \bibfield  {author} {\bibinfo {author} {\bibfnamefont {T.~A.}\ \bibnamefont
  {Maier}}, \bibinfo {author} {\bibfnamefont {M.~S.}\ \bibnamefont {Jarrell}},
  \ and\ \bibinfo {author} {\bibfnamefont {D.~J.}\ \bibnamefont {Scalapino}},\
  }\href@noop {} {\bibfield  {journal} {\bibinfo  {journal} {Physical Review
  Letters}\ }\textbf {\bibinfo {volume} {96}},\ \bibinfo {pages} {047005}
  (\bibinfo {year} {2006})}\BibitemShut {NoStop}%
\bibitem [{\citenamefont {Staar}\ \emph {et~al.}(2014)\citenamefont {Staar},
  \citenamefont {Maier},\ and\ \citenamefont {Schulthess}}]{Staar:2014jy}%
  \BibitemOpen
  \bibfield  {author} {\bibinfo {author} {\bibfnamefont {P.}~\bibnamefont
  {Staar}}, \bibinfo {author} {\bibfnamefont {T.}~\bibnamefont {Maier}}, \ and\
  \bibinfo {author} {\bibfnamefont {T.~C.}\ \bibnamefont {Schulthess}},\
  }\href@noop {} {\bibfield  {journal} {\bibinfo  {journal} {Physical Review
  B}\ }\textbf {\bibinfo {volume} {89}},\ \bibinfo {pages} {195133} (\bibinfo
  {year} {2014})}\BibitemShut {NoStop}%
\end{thebibliography}%


%
\end{document}